\documentclass[12pt]{article}
\usepackage{graphicx}
\begin{document} 

\title{Phase diagram in Bonabeau social hierarchy model with 
individually different abilities}

\author{Christian Schulze and Dietrich Stauffer\\
Institute for Theoretical Physics, Cologne University\\D-50923 K\"oln, Euroland}

\maketitle
\centerline{e-mail: stauffer@thp.uni-koeln.de}

\bigskip
Abstract: 
The 1995 model of Bonabeau et al is generalized by giving each
individual a different ability to win or lose a fight. We also
introduce different groups such that the losers of fights between
different groups are eliminated. The overall phase diagram for 
the first-order transition between egalitarian and hierarchical
societies does not change much by these generalizations.

\bigskip


\bigskip

\section{Introduction}

The model of Bonabeau et al \cite{bonabeau} described how 
a difference between powerful and powerless levels of society
can self-organize out of randomness and the memory of past 
power fights. With suitable modifications \cite{stauffer} a
first-order phase transition was found: at low concentrations
the society remained egalitarian, while at high concentrations
some individuals became "more equal" than others. These 
inequalities are measured by the probability to win the fights
arising whenever one individual wants to move onto the place 
occupied by another; initially we have an egalitarian society
where everybody has the winning probability 1/2. As in some
biological species \cite{cray}, the individuals keep a 
memory of past fights such that past victories
re-inforce the probability to win again. The present 
note makes the individuals unequal from the beginning and checks
how this modification changes the results.

\section{Old Model}

A $L \times L$ square lattice is filled randomly with $N = pL^2$
individuals such that no two or more share one site. Then they 
diffuse randomly to nearest-neighbour sites. If one individual
$i$ wants to occupy the site on which already another individual
$k$ sits, a fight breaks out which is won by one of the two, who
then moves into the contested site; the loser instead moves into
the site previously occupied by the winner. The probability for
$i$ to win is $$ q = 1/[1 + \exp(\sigma \cdot (h(k)-h(i))]$$ 
where $h(i)$ is the
number of past victories minus the number of past losses of $i$
except that at each time step all $h$ are diminished by a factor
$1 - f$; this memory factor $f$, which is the same for all 
individuals, indicates how fast past events are forgotten. 
The inequality $\sigma$ is the standard deviation in the 
probabilities $q$: $$ \sigma^2 = <q^2> - <q>^2 \quad ;$$ during the 
first 10 iterations we set $\sigma = 1$. With this feedback
through $\sigma$ a sharp first-order phase transition 
(discontinuity) was found
\cite{stauffer} near $p = 0.32$ for $f = 0.1$: For higher 
concentrations $p$ a non-zero social inequality $\sigma$ was
obtained while for lower $p$ this $\sigma$ vanished to zero.

\begin{figure}[hbt]
\begin{center}
\includegraphics[angle=-90,scale=0.5]{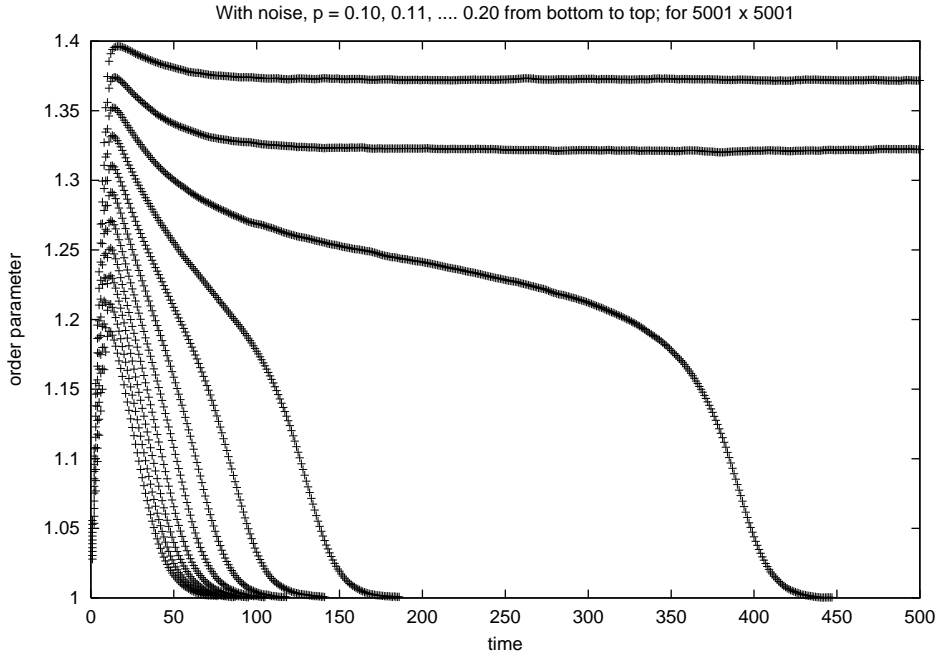}
\end{center}
\caption{Search for phase transition with permanent random abilities, at 
$f=0.1$. For the more complicated models 
discussed later the order parameter did not always go to unity if $\sigma 
\rightarrow 0$, and thus $\sigma$ which gave similar figures was used as a 
criterion. 
}
\end{figure}

\begin{figure}[hbt]
\begin{center}
\includegraphics[angle=-90,scale=0.5]{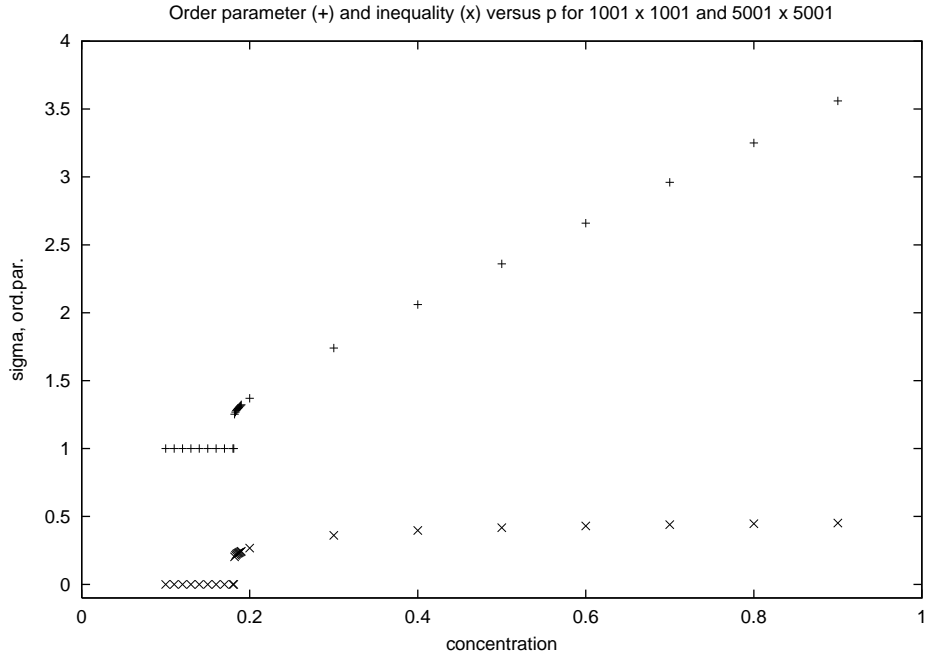}
\end{center}
\caption{Stationary values for order parameter and inequality, from simulations 
like in Fig.1, using medium and large lattices. 
}
\end{figure}

\begin{figure}[hbt]
\begin{center}
\includegraphics[angle=-90,scale=0.33]{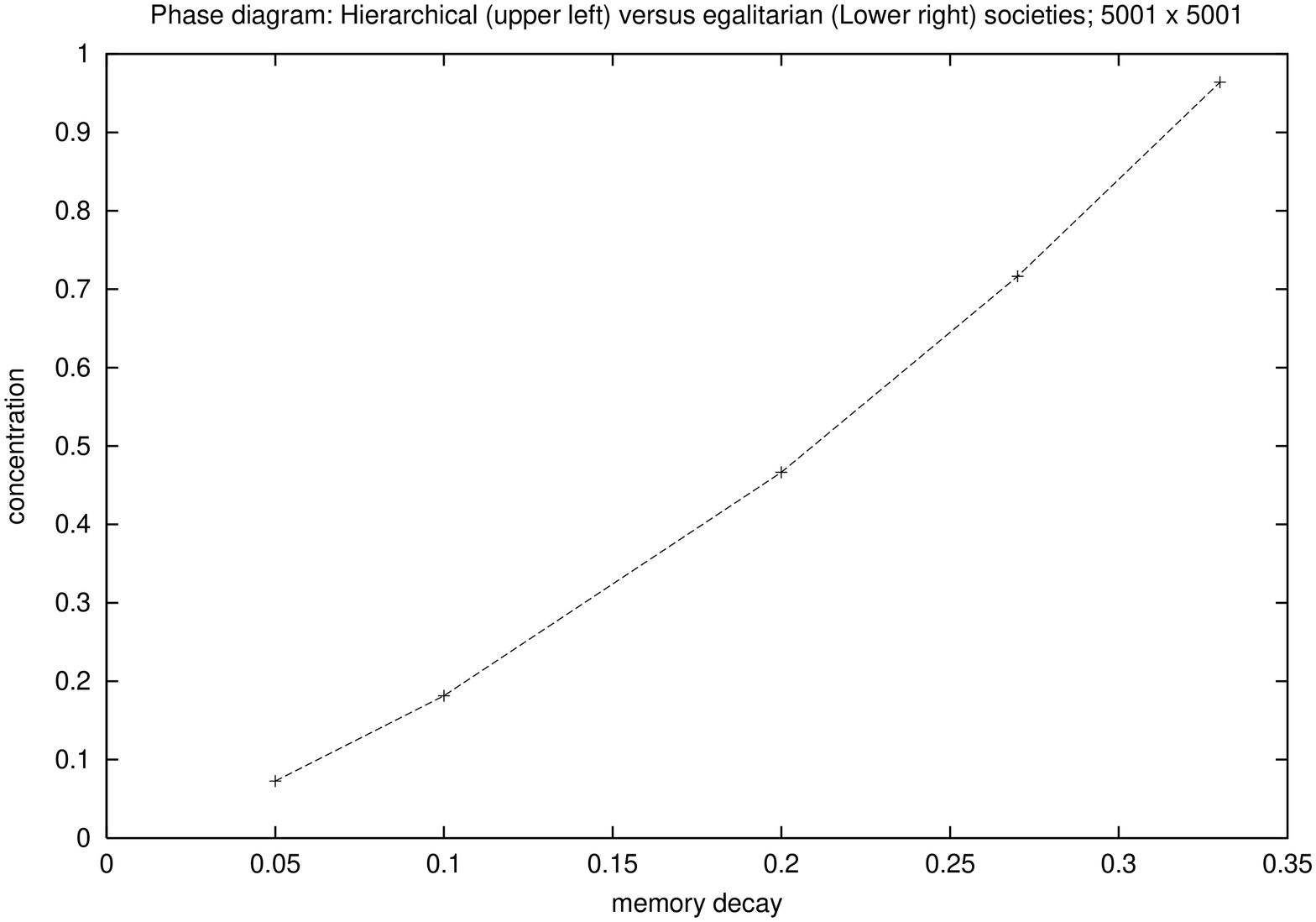}
\includegraphics[angle=-90,scale=0.33]{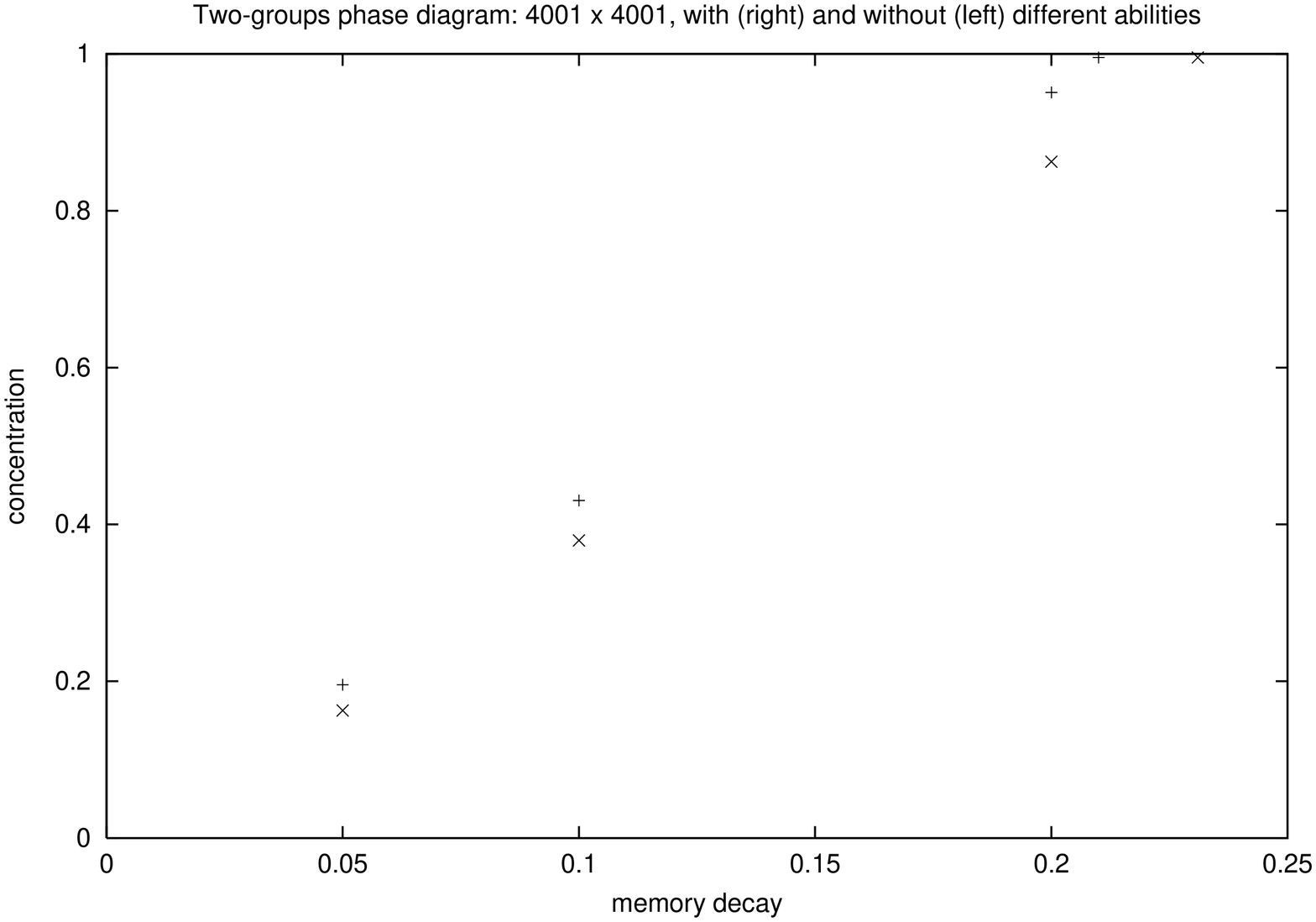}
\end{center}
\caption{Phase diagram with one group (part a) and two groups (part b); in
the latter case also the results without different abilities are shown.}
\end{figure}

\begin{figure}[hbt]
\begin{center}
\includegraphics[angle=-90,scale=0.5]{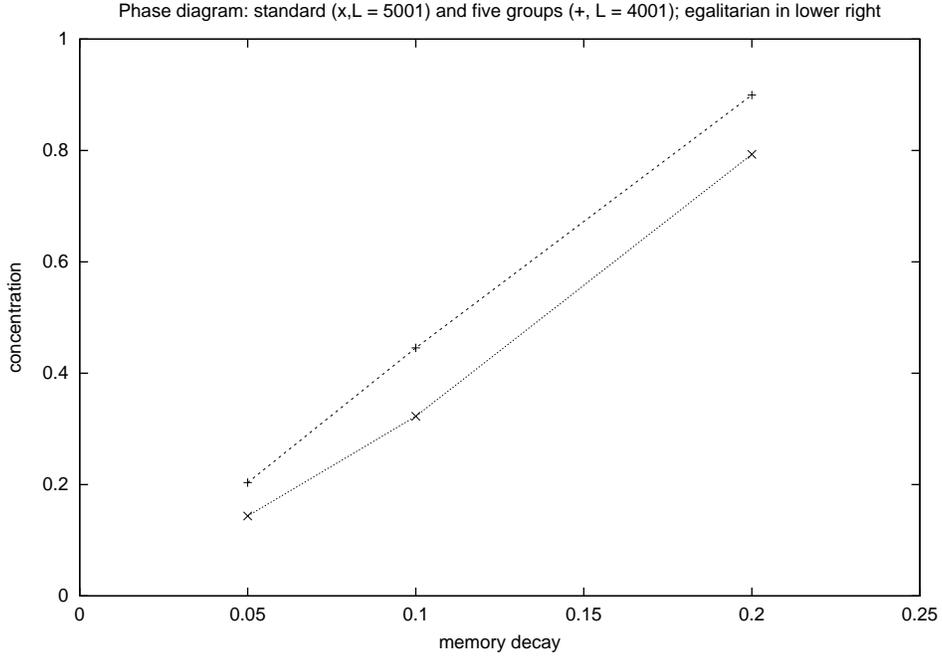}
\end{center}
\caption{Phase diagram for five groups (+); the x indicate the standard model
without different groups and without different abilities.}
\end{figure}

\begin{figure}[hbt]
\begin{center}
\includegraphics[angle=-90,scale=0.5]{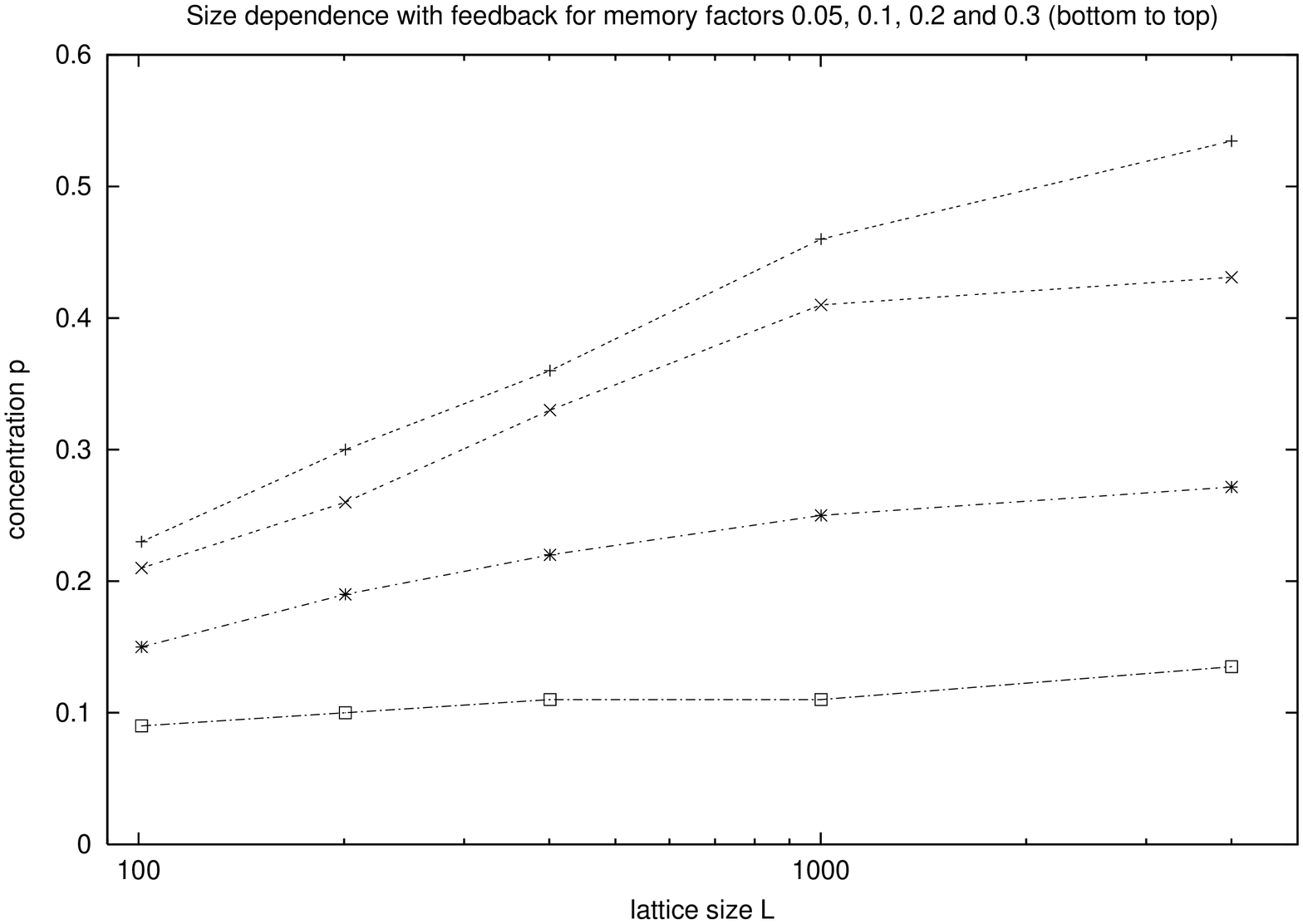}
\end{center}
\caption{Size effects in the feedback model, indicating a roughly logarithmic
increase of the transition $p$ with $L$.}
\end{figure}

\section{New Models}

First we gave each individual $i$ different histories $h(i)$ to start with;
then initially the preferred individuals for low concentrations
could enlarge their advantage but after some time they lost it
and no correlations were seen between initial and final $h$
(not shown).

Second, we gave each individual different abilities to win a 
fight
by replacing $h(i)$ with $h(i) + a(i)z$ where $z$ is a random 
number between 0 and 1, evaluated for each fight and each $i$
again and again, while the ability $a(i)$ stays with each
individual $i$ forever and is initially distributed randomly
between $-5$ and +5. (Setting $z = 1$ does not change much.)

Third, we divided the whole population into $G$ groups, 
initially equally large. 
If two individuals of the same group fight, the 
above rules apply; if two individuals of different groups 
fight, the loser vanishes and is replaced by a replica of the 
winner with the same $h$ and $a$; thus the winner is duplicated
and occupies both sites \cite{ispolatov}.

Fourth, a feedback \cite{bornholdt} for the special case $G = 2$
is introduced such that the majority wins more easily than the 
minority. Thus the above $h(i) + a(i)z$ is replaced by
$h(i)+[a(i) + 10 m]z$ where $m$ with $-1 < m < +1$ is the 
magnetization, i.e. the normalized difference 
$(N_1-N_2)/(N_1+N_2)$ between the numbers $N_1, \; N_2$
of individuals in the two groups. 

The similarity between the final power $h(i)+a(i)$ and its initial value
$a(i)$ is given by the order parameter $$\Psi \propto \sum_i [h(i)+a(i)]a(i)$$
which we normalize to unity at the beginning. 
 
\section{Results}

Usually we made up to 5000 iterations with up to about $5000 \times 5000$ 
sites. For the second model, 
Fig.1 shows how the order parameter initially increases due to re-inforcement 
of individuals with higher abilities. However, if the concentration 
is low such that fights occur seldomly, then the positive effects of these 
abilities are mostly forgotten until the next fight, and everybody has the 
same chance of 1/2 to win or lose. This happens for the 8 leftmost curves, $p =
0.10, 0.11, \dots, 0.17$, in Fig.1. For $p = 0.18$ the system needs a long time
to find this egalitarian state. For $p = 0.19$ and 0.20 we see two plateaus 
in the later times, with no indication of a decay. With additional runs at
$p = 0.181, 0.182, \dots$ and 5000 iterations we found a stable order paraneter
$\Psi > 1$ and inequality $\sigma > 0$  only at $p \ge 0.182$. Fig.2 shows the
long-time values of $\Psi$ and $\sigma$, indicating a jump (first-order 
transition) of both quantities near $p = 0.182$. All these simulations were 
made with the traditional value $f = 0.1$. 

Varying the memory factor $f$ we find the phase diagram of Fig.3a; for $f > 1/3$ 
the model always leads to egalitarian results since past fights are forgotten
too fast. Fig.3b shows the analogous phase diagram in the third model with two 
groups, also for the case without different abilities. 

With five groups, Fig.4 shows a behaviour similar to one or two groups, Fig.3.
We also give here as the lower data the results for the standard model 
\cite{stauffer} without different abilities and only one group; again, the 
behaviour seems similar. Thus for all models except the fourth one the results
are similar: The transition line starts at the origin and increases roughly
linearly to a fully occupied lattice, $p = 1$, at a memory decay factor between
1/5 and 1/3; for larger $f$ (shorter memory) the society remains egalitarian.

In the fourth model with feedback the concept of a transition is less clear. 
The larger the lattice is, the smaller is the normalized fluctuating 
magnetization $m$ appearing in $h(i)+[a(i) + 10 m]z$ and the higher is the
critical concentration at which the egalitarian society is destroyed. The
earlier transition points were determined by gradually increasing $p$, Fig.1;
in the feedback version, however, it happened that when at some $p$ the 
inequality $\sigma$ went to zero, at some slightly higher $p$ it became nonzero,
and then zero again for still higher $p$. In that case, for large $L = 4001$ and one
sample only, some intermediate $p$ value was taken as transition point. For
smaller $101 \le L \le 1001$, hundred samples were simulated and the transition
point defined as the concentration $p$ where about 50 samples ended with 
$\sigma = 0$ and the others with nonzero $\sigma$. Fig.5 shows an increase
of the transition $p$ roughly logarithmic in system size; and for infinite
system size the whole feedback effect as simulated here would vanish and the
results agree with those of Fig.3b. 

We thank D. Sornette for suggesting to put individual differences
into the Bonabeau model.

\end{document}